\documentstyle[lprocl,11pt,epsf,epsfig]{article}

\def\NPB{{\em Nucl. Phys.} B}
\def\PLB{{\em Phys. Lett.}  B}
\def\PRL{\em Phys. Rev. Lett.}
\def\PRD{{\em Phys. Rev.} D}
\def\ZPC{{\em Z. Phys.} C}

\def\be{\begin{equation}}
\def\ee{\end{equation}}
\def\bea{\begin{eqnarray}}
\def\eea{\end{eqnarray}}

\def\beq{\begin{equation}}
\def\eeq{\end{equation}}
\def\bea{\begin{eqnarray}} 
\def\eea{\end{eqnarray}}
\def\bq{\begin{quote}} 
\def\eq{\end{quote}}

\def\PR{{\it Phys.Rev.} } 
\def\PRL{{\it Phys.Rev.Lett.} }

\def\gappeq{\mathrel{\rlap {\raise.5ex\hbox{$>$}}
{\lower.5ex\hbox{$\sim$}}}}

\def\lappeq{\mathrel{\rlap{\raise.5ex\hbox{$<$}}
{\lower.5ex\hbox{$\sim$}}}}

\begin{document}

\title{
\vskip -1cm
\begin{flushright} 
{\rm CERN-TH.97-278 \\ 
October 1997} 
\end{flushright}
\vskip +1cm
THE STATUS OF THE STANDARD MODEL}

\author{G. Altarelli}

\address{Theoretical Physics Division, CERN\\
 1211 Geneva, 23, Switzerland\\
 Universit\`a di Roma Tre, Rome, Italy\\
 E-mail: Guido.Altarelli@cern.ch}   




\maketitle\abstracts{1. Introduction\\
2. Status of the Data \\
3. Precision Electroweak Data and
the Standard Model\\
4. A More General Analysis of Electroweak Data\\
5. The Case for Physics beyond the Standard Model\\
6. The Search for the
Higgs\\
7. New Physics at HERA?\\
8. Conclusion  \\}


\section{ Introduction}

In recent years new powerful tests of the Standard Model (SM) have been
performed mainly at LEP but also at SLC and at the Tevatron. The running
of LEP1 was terminated in 1995 and close-to-final results of the data
analysis are now available~\cite{tim}$^,$\cite{ew}. The experiments at the
$Z_0$ resonance have enormously improved the accuracy in the electroweak
neutral current sector. The LEP2 programme is in progress. I went back to
my rapporteur talk at the Stanford Conference in August 1989~\cite{sta} and I~found the following best values quoted there for some of
the key quantities of interest for the Standard Model (SM)
 phenomenology: $m_Z$ = 91120(160) MeV; $m_t$ = 130(50) GeV;
$\sin^2\theta_{eff}$ = 0.23300(230); $m_H\gappeq$ a few GeV  and
$\alpha_s(m_Z)$ = 0.110(10).  Now, after seven years of experimental and
theoretical work (in particular with $\sim16$ million Z events analysed
 altogether by the four LEP experiments) the corresponding numbers,
 as quoted at this Symposium, are: $m_Z$ = 91186.7(2.0) MeV; $m_t$ =
175.6(5.5) GeV;
$\sin^2\theta_{eff}$ = 0.23152(23), $m_H\gappeq 77~GeV$ and 
$\alpha_s(m_Z)$ = 0.119(3). Thus the progress is quite evident. The top
quark has been at last found and  the errors on $m_Z$ and
$\sin^2\theta_{eff}$ went down by two and one orders of magnitude
respectively. The validity of the SM has been confirmed to a level that we
can say was unexpected at the beginning. In the present data there is no
significant evidence for departures from the SM, no convincing hint of new
physics (also including the first  results from LEP2)~\cite{dio}. The
impressive success of the SM poses strong limitations on the possible
forms of  new physics. Favoured are models of the Higgs sector and of new
physics that preserve the SM structure  and only very delicately improve
it, as is the case for fundamental Higgs(es) and Supersymmetry.
Disfavoured are models with a nearby strong non perturbative regime that 
almost inevitably would affect the radiative corrections, as for composite
Higgs(es) or technicolor and its variants. 

\section{ Status of the Data}

The relevant electro-weak data together with their SM values are presented
in table 1~\cite{tim}$^-$\cite{ew}.  The SM predictions correspond to a fit
of all the available data (including the directly measured values of $m_t$
and
$m_W$) in terms of $m_t$, $m_H$ and $\alpha_s(m_Z)$, described later in
sect.3, table 4.

Other important derived quantities are, for example, $N_\nu$ the number of
light neutrinos, obtained from the invisible width: $N_\nu=2.993(11)$,
which shows that only three fermion generations exist with $m_\nu
<45~GeV$, or the leptonic width $\Gamma_l$, averaged over e, $\mu$ and
$\tau$:
$\Gamma_l= 83.91(10) MeV$.   

For indicative purposes, in table 1 the "pulls" are also shown, defined
as: pull = (data point- fit value)/(error on data point).  At a glance we
see that the agreement with the SM is quite good. The distribution of the
pulls is statistically normal. The presence of a few $\sim2\sigma$
deviations is what is to be expected. However it is maybe worthwhile to
give a closer look at these small discrepancies.

Perhaps the most annoying feature of the data is the persistent difference
between the values of
$\sin^2\theta_{eff}$ measured at LEP and at SLC. The value of
$\sin^2\theta_{eff}$ is obtained from a set of combined asymmetries. From
asymmetries one derives the ratio $x=g_V^l/g_A^l$ of the vector and axial
vector couplings of the $Z_0$, averaged over the charged leptons. In turn
$\sin^2\theta_{eff}$ is defined by $x=1-4\sin^2\theta_{eff}$. SLD obtains
x from the single measurement of
$A_{LR}$, the left-right asymmetry, which requires longitudinally
polarized beams. The distribution of the present measurements of
$\sin^2\theta_{eff}$ is shown in fig.1. The LEP average,
$\sin^2\theta_{eff}=0.23199(28)$, differs by
$2.9\sigma$ from the SLD value
$\sin^2\theta_{eff}=0.23055(41)$. The most precise individual measurement
at LEP is from $A^{FB}_b$: the combined LEP error on this quantity is
about the same as the SLD error, but the two values are $3.1\sigma$'s
away. One might attribute this to the fact that the b measurement is more
delicate and affected by a complicated systematics. In fact one notices
from fig.1 that the value  obtained at LEP from $A^{FB}_l$, the average
for l=e, $\mu$ and $\tau$, is somewhat low (indeed quite in agreement with
the SLD value). However the statement that LEP and SLD agree on leptons
while they only disagree when the b quark is considered is not quite
right. First, the value of
$A_e$, a quantity essentially identical to
$A_{LR}$, measured at LEP from the angular distribution of the $\tau$
polarization, differs by
$1.8\sigma$ from the SLD value. Second, the low value of
$\sin^2\theta_{eff}$ found at LEP from
$A^{FB}_l$  turns out to be entirely due to the $\tau$ lepton channel
which leads to a central value different than that of e and
$\mu$~\cite{ew}.The e and $\mu$ asymmetries, which are experimentally
simpler, are perfectly on top of the SM fit. Suppose we take only e and
$\mu$ asymmetries at LEP and disregard the b and $\tau$ measurements: the
LEP average becomes $\sin^2\theta_{eff}=0.23184(55)$, which is still
$1.9\sigma$ away from the SLD value.

\begin{figure}
\hglue 2.5cm
\epsfig{figure=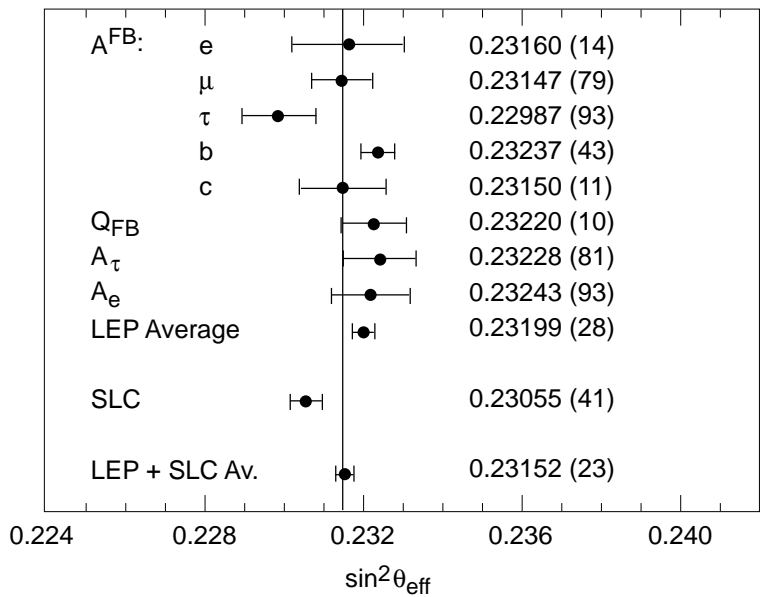,width=10cm}
\caption[]{The collected measurements of $\sin^2\theta_{eff}$. The resulting value
for the $\chi^2$ is given by
$\chi^2/d.o.f=1.87$. As a consequence the error on the average is enlarged
in the text by a factor $\sqrt{1.87}$ with respect to the formal average
shown here.}
\end{figure} 

In conclusion, it is difficult to find a simple explanation for the
SLD-LEP discrepancy on
$\sin^2\theta_{eff}$. In view of this, the error on the nominal SLD-LEP
average,
$\sin^2\theta_{eff}=0.23152(23)$, should perhaps be enlarged, for example,
by a factor
$S=\sqrt{\chi^2/N_{df}}\sim1.4$, according to the recipe adopted by the
Particle Data Group in such cases. Accordingly, in the following we will
often use the average
\beq 
\sin^2\theta_{eff}=0.23152\pm0.00032
 \label{8}
  \eeq
         Thus the LEP-SLC
discrepancy results in an effective limitation of the experimental
precision on
$\sin^2\theta_{eff}$. The data-taking by the SLD experiment is still in
progress and also at LEP seizable improvements on
$A_{\tau}$ and $A^{FB}_b$ are foreseen as soon as the corresponding
analyses will be completed. We hope to see the difference to decrease or
to be understood.

From the above discussion one may wonder if there is evidence for
something special in the $\tau$ channel, or equivalently if lepton
universality is really supported by the data. Indeed this is the case: the
hint of a difference in $A^{FB}_\tau$ with respect to the corresponding e
and
$\mu$ asymmetries is not confirmed by the measurements of  $A_\tau$ and 
$\Gamma_\tau$ which appear normal~\cite{tim}$^,$\cite{ew}$^,$\cite{li}. In principle the fact that an anomaly
shows up in $A^{FB}_\tau$  and not in
$A_\tau$ and 
$\Gamma_\tau$ is not unconceivable because the FB lepton asymmetries are
very small and very precisely measured. For example, the extraction of
$A^{FB}_\tau$ from the data on the angular distribution of $\tau$'s could
be biased if the imaginary part of the continuum was altered by some non
universal new physics effect~\cite{car}. But a more trivial experimental
problem is at the moment quite plausible.

\begin{table}[t]
\caption{}
\vspace{0.2cm}
\begin{center}
\footnotesize
\begin{tabular}{|l|l|l|l|}
\hline Quantity&Data (August '97)       & Standard Model Fit & Pull\\
\hline
$m_Z$ (GeV)&91.1867(20) &91.1866 &~~0.0\\
$\Gamma_Z$ (GeV)        &2.4948(25) & 2.4966 & $-0.7 $\\
$\sigma_h$ (nb) &41.486(53)     &41.467 & ~~0.4\\
$R_h$   &20.775(27)     &20.756 & ~~0.7\\
$R_b$ &0.2170(9)       &0.2158 & ~~1.4\\
$R_c$&  0.1734(48)&     0.1723 & $-0.1$ \\
$A^l_{FB}$&  0.0171(10) &0.0162 & ~~0.9 \\
$A_\tau$ &      0.1411(64)      &0.1470 & $-0.9$ \\
$A_e$   &0.1399(73) &0.1470& $-1.0$\\
$A^b_{FB}$ &    0.0983(24) &0.1031 & $-2.0$ \\
$A^c_{FB}$&     0.0739(48)      &0.0736 & ~~0.0\\
$A_b$ (SLD direct)   & 0.900(50) &0.935 & $-0.7$\\ 
$A_c$ (SLD direct)  &  0.650(58) &0.668 & $-0.3$\\ 
$\sin^2\theta_{eff}({\rm\hbox{LEP-combined}})$ & 0.23199(28) & 0.23152 &
~~1.7\\
$A_{LR}\rightarrow  \sin^2\theta_{eff}$& 0.23055(41) &  0.23152 & $-2.4$
\\
$m_W$ (GeV) (LEP2+p$\bar p$) & 80.43(8)       &80.375& ~~0.7\\
$1-\frac{m^2_W}{m^2_Z}$ ($\nu$N) &  0.2254(37) &0.2231 & $~~0.6$\\
$Q_W$ (Atomic PV in Cs) &  -72.11(93) &-73.20 & $~~1.2$\\
$m_t$ (GeV)     &175.6(5.5) &173.1 & ~~0.4\\
\hline
\end{tabular}
\end{center}
\end{table}

A similar question can be asked for the b couplings. We have
seen that the measured value of $A^{FB}_b$ is about $2\sigma$'s below the
SM fit. At the same time $R_b$ which used to show a major discrepancy is
now only about $1.4\sigma$'s away from the SM fit (as a result of the more
sophisticated second generation experimental techniques).       It is often
stated that there is a $-2.5\sigma$ deviation on  the measured value of
$A_b$ vs the SM expectation~\cite{tim}$^,$\cite{ew}. But in fact that 
depends on how the data are combined. In our opinion one should rather
talk of a $-1.8\sigma$ effect. Let us discuss this point in detail.
$A_b$ can be measured directly at SLC by taking advantage of the beam
longitudinal polarization. At LEP one measures
$A^{FB}_b$    = 3/4 $A_eA_b$. One can then derive $A_b$ by inserting a
value for $A_e$. The question is what to use for $A_e$: the LEP value
obtained, using lepton universality, from the measurements of $A^{FB}_l$,
$A_\tau$, $A_e$: $A_e$ = 0.1461(33), or the combination of LEP and SLD
etc. The LEP electroweak working group adopts for
$A_e$ the SLD+LEP average value which also includes $A_{LR}$ from SLD:
$A_e$ = 0.1505(23). This procedure leads to a $-2.5\sigma$ deviation.
However, in this case, the well known $\sim 2\sigma$ discrepancy of
$A_{LR}$ with respect to $A_e$ measured at LEP and also to the SM fit,
which is not related to the b couplings, further contributes to inflate
the number of $\sigma$'s. Since we are here concerned with the b couplings
it is perhaps wiser to obtain $A_b$ from LEP by using the SM value for
$A_e$ (that is the pull-zero value of table 1):
$A^{SM}_e$   = 0.1467(16). With the value of $A_b$ derived in this way
from LEP we finally obtain \beq
 A_b = 0.895\pm0.022~~~~~(\rm{LEP+SLD, A_e=A^{SM}_e: -1.8}) \label{7} \eeq
In the SM $A_b$ is so close to 1 because the b quark is almost purely
left-handed. $A_b$ only depends on the ratio $r=(g_R/g_L)^2$ which in the
SM is small: $r\sim 0.033$. To adequately decrease $A_b$ from its SM value
one must increase r by a factor of about 1.6, which appears large for a
new physics effect. Also such a large change in x must be compensated by
decreasing $g_L^2$ by a small but fine-tuned amount in order to
counterbalance the correponding large positive shift in $R_b$. In view of
this the most likely way out is that $A^{FB}_b$ and
$A_b$ have been a bit underestimated at LEP and actually there is no
anomaly in the b couplings. Then the LEP value of $\sin^2\theta_{eff}$
would slightly move towards the SLD value, but, as explained above, by far
not enough to remove the SLD-LEP discrepancy (for example, if the LEP
average for $\sin^2\theta_{eff}$ is computed by moving the central value
of $A^{FB}_b$ to the pull-zero value in Table 1 with the same figure for
the error, one finds $\sin^2\theta_{eff}=0.23162(28)$, a value still
$2.2\sigma$'s away from SLD).

\section{ Precision Electroweak Data and the Standard Model}

        For the analysis of electroweak data in the SM one starts from the input
parameters: some of them,
$\alpha$, $G_F$ and $m_Z$, are very well measured, some other ones,
$m_{f_{light}}$, $m_t$ and
$\alpha_s(m_Z)$  are only approximately determined while $m_H$ is largely
unknown. With respect to
$m_t$ the situation has much improved since the CDF/D0 direct measurement
of the top quark mass~\cite{gir}. From the input parameters one computes the radiative
corrections~\cite{radcorr} to a sufficient precision to match the
experimental capabilities. Then compares the theoretical predictions and
the data for the numerous observables which have been measured, checks the
consistency of the theory and derives constraints on $m_t$,
$\alpha_s(m_Z)$ and hopefully also on $m_H$. 

        Some comments on the least known of the input parameters are now in
order. The only practically relevant terms where precise values of the
light quark masses, $m_{f_{light}}$, are needed are those related to the
hadronic contribution to the photon vacuum polarization diagrams, that
determine
$\alpha(m_Z)$. This correction is of order 6$\%$, much larger than the
accuracy of a few permil of the precision tests. For some direct experimental
evidence on the running of $\alpha(Q^2)$, see ref. \cite{mk}.  Fortunately, one can use
the actual data to in principle solve the related ambiguity. But the
leftover uncertainty is still one of the main sources of theoretical
error. DA$\phi$NE can in the near future reduce somewhat this error (and also
that on the related hadronic contribution to the muon $g-2$ of relevance for
the ongoing Brookhaven measurement~\cite{lmb}).  In recent years there has
been a lot of activity on this subject
and a number of independent new estimates of $\alpha(m_Z)$  have appeared
in the literature~\cite{alfaQED},(see also~\cite{piet}). A consensus has been established and the value used at
present is
\beq
\alpha(m_Z)^{-1}=128.90\pm0.09 
\label{8ab} 
\eeq  

        As for the strong coupling $\alpha_s(m_Z)$ the world average central
value is by now quite stable. The error is going down because the
dispersion among the different measurements is much smaller in the most
recent set of data. The most important determinations of $\alpha_s(m_Z)$
are summarised in table 2~\cite{cat}. For all entries, the main sources of
error are the theoretical ambiguities which are larger than the
experimental errors. The only exception is the measurement from the
electroweak precision tests, but only if one assumes that the SM
electroweak sector is correct. Our personal views on the theoretical
errors are reflected in the table 2. The error on the final average is
taken by all authors between
$\pm$0.003 and
$\pm$0.005 depending on how conservative one is. Thus in the following our
reference value will be
 \beq 
\alpha_s(m_Z) = 0.119\pm0.004
 \label{9}
  \eeq

\begin{table}[t]
\caption{Measurements of $\alpha_s(m_Z)$. In
parenthesis we indicate if the dominant source of errors is theoretical or
experimental. For theoretical ambiguities our personal figure of merit is
given.}
\vspace{0.2cm}
\begin{center}
\footnotesize
\begin{tabular}{|l|ll|}
\hline Measurements & \multicolumn{2}{c|}{$\alpha_s(m_Z)$}\\
\hline
$R_{\tau}$ & 0.122 $\pm$ 0.006 & (Th)\\ Deep Inelastic Scattering & 0.116
$\pm$ 0.005 & (Th)\\
$Y_{\rm decay}$ & 0.112 $\pm$ 0.010 & (Th)\\ Lattice QCD & 0.117 $\pm$
0.007 & (Th)\\
$Re^+e^-(\sqrt s < 62~{\rm GeV}$) & 0.124 $\pm$ 0.021 & (Exp)\\
Fragmentation functions in $e^+e^-$ & 0.124 $\pm$ 0.012 & (Th)\\ Jets in
$e^+e^-$ at and below the $Z$ & 0.121 $\pm$ 0.008 & (Th)\\
$Z$ line shape (Assuming SM) & 0.120 $\pm$ 0.004 & (Exp)\\
\hline
\end{tabular}
\end{center}
\end{table}
Finally a few words on the current status of the direct measurement of
$m_t$. The present combined CDF/D0 result is~\cite{gir}
\beq  m_t = 175.6\pm 5.5~GeV
 \label{10}
  \eeq
   The error is so small by now that one is
approaching a level where a more careful investigation of the effects of
colour rearrangement on the determination of $m_t$ is needed. One wants to
determine the top quark mass, defined as the invariant mass of its decay
products (i.e. b+W+ gluons +
$\gamma$'s). However, due to the need of colour rearrangement, the top
quark and its decay products cannot be really isolated from the rest of
the event. Some smearing of the mass distribution is induced by this
colour crosstalk which involves the decay products of the top, those of
the antitop and also the fragments of the incoming (anti)protons. A
reliable quantitative computation of the smearing effect on the $m_t$ 
determination is difficult because of the importance of non perturbative
effects. An induced error of the order of a few GeV on $m_t$ is reasonably
expected. Thus further progress on the $m_t$ determination demands
tackling this problem in more depth. 

        In order to appreciate the relative importance of the different sources
of theoretical errors for precision tests of the SM, I report in table 3 
a comparison for the most relevant observables, evaluated using refs.~\cite{radcorr}$^,$\cite{radcorr2}.        What is important to stress is that the
ambiguity from $m_t$, once by far the largest one, is by now smaller than
the error from $m_H$. We also see from table 3 that the error from
$\Delta\alpha(m_Z)$ is expecially important for $\sin^2\theta_{eff}$  and,
to a lesser extent, is also sizeable for
$\Gamma_Z$ and $\epsilon_3$.  
\begin{table}[t]
\caption{Measurements of $\alpha_s(m_Z)$. In
parenthesis we indicate if the dominant source of errors is theoretical or
experimental. For theoretical ambiguities our personal figure of merit is
given.}
\vspace{0.2cm}
\begin{center}
\footnotesize
\begin{tabular}{|l|l|l|l|l|l|l|}
\hline Parameter& $\Delta^{exp}_{now}$ & $\Delta \alpha^{-1}$ &
$\Delta_{th}$ &
$\Delta m_t$ & $\Delta m_H$ & $\Delta \alpha_s$ \\
\hline
$\Gamma_Z$ (MeV) & $\pm$2.5 & $\pm$0.7 & $\pm$0.8 & $\pm$1.4 & $\pm$4.6 &
$\pm$1.7 \\
$\sigma_h$ (pb) & 53 & 1 & 4.3 & 3.3 & 4 & 17\\
$R_h \cdot 10^3$ & 27 & 4.3 & 5 & 2 & 13.5 & 20 \\
$\Gamma_l$ (keV) & 100 & 11 & 15 & 55 & 120 & 3.5\\
$A^l_{FB}\cdot 10^4$ & 10 & 4.2 & 1.3 & 3.3 & 13 & 0.18 \\
$\sin^2\theta\cdot 10^4$ & $\sim$3.2 & 2.3 & 0.8 & 1.9 & 7.5 & 0.1\\
$m_W$~(MeV) & 80 & 12 & 9 & 37 & 100& 2.2 \\
$R_b \cdot 10^4$ & 9 & 0.1 & 1 & 2.1 & 0.25 & 0\\
$\epsilon_1\cdot 10^3$ & 1.2 & & $\sim$0.1 & & & 0.2\\
$\epsilon_3\cdot 10^3$ & 1.4 & 0.5 & $\sim$0.1 & & & 0.12\\
$\epsilon_b\cdot 10^3$ & 2.1 & & $\sim$0.1 & & & 1\\
\hline
\end{tabular}
\end{center}
\end{table}

The most important recent advance in the theory of radiative corrections
is the calculation of the 
$o(g^4m^2_t/m^2_W)$ terms in $\sin^2\theta_{eff}$ and $m_W$ (not yet in
$\delta\rho$)~\cite{deg}. The result implies a small but visible
correction to the predicted values but expecially a seizable decrease of
the ambiguity from scheme dependence (a typical effect of truncation).  

        We now discuss fitting the data in the SM. Similar studies based on older
sets of data are found in refs.~\cite{fits}. As the mass of the top quark
is finally rather precisely known from CDF and D0 one must distinguish two
different types of fits. In one type one wants to answer the question: is
$m_t$ from radiative corrections in agreement with the direct measurement
at the Tevatron? For answering this interesting but somewhat limited
question, one must clearly exclude the CDF/D0 measurement of $m_t$ from
the input set of data. Fitting all other data in terms of
$m_t$,
$m_H$ and
$\alpha_s(m_Z)$ one finds the results shown in the third column of table 4~\cite{ew}. Other similar fits where also $m_W$ direct data are left out
are shown.
\begin{table}[t]
\caption{}
\vspace{0.2cm}
\begin{center}
\footnotesize
\begin{tabular}{|l|l|l|l|l|}
\hline Parameter & LEP(incl.$m_W$) &All but $m_W$, $m_t$ &All but  $m_t$ &
All Data\\
\hline
$m_t$ (GeV) & 158$+14-11$ & 157$+10-9$ & 161$+10-8$ & $173.1\pm5.4$\\
$m_H$ (GeV) & 83$+168-49$ & 41$+64-21$ & 42$+75-23$ & 115$+116-66$\\
$log[m_H(GeV)]$ & 1.92$+0.48-0.39$ &  1.62$+0.41-0.31$ &  1.63$+0.44-0.33$
& 2.06$+0.30-0.37$\\
$\alpha_s(m_Z)$ & $0.121\pm0.003$ & $0.120\pm0.003$ & $0.120\pm0.003$ &
$0.120\pm0.003$ \\
$\chi^2/dof$ & 8/9 & 14/12 & 16/14 & 17/15\\
\hline
\end{tabular}
\end{center}
\end{table}

The extracted value of $m_t$ is typically a bit too low. There
is a strong correlation between $m_t$ and $m_H$.
$\sin^2\theta_{eff}$ and $m_W$~\cite{kim} drive the fit to small values of
$m_H$. Then, at small $m_H$ the widths, drive the fit to small $m_t$. In
this context it is important to remark that fixing
$m_H$ at 300 GeV, as was often done in the past, is by now completely
obsolete, because it introduces too strong a bias on the fitted value of
$m_t$. The change induced on the fitted value of $m_t$ when moving $m_H$
from 300 to 65 or 1000 GeV is in fact larger than the error on the direct
measurement of $m_t$. 

        In a more general type of fit, e.g. for determining the overall
consistency of the SM or the best present estimate for some quantity, say
$m_W$, one should of course not ignore the existing direct determinations
of $m_t$ and $m_W$. Then, from all the available data,  by fitting
$m_t$, $m_H$ and $\alpha_s(m_Z)$ one finds the values shown in the last
column of table 4. This is the fit also referred to in table 1. The
corresponding fitted values of $\sin^2\theta_{eff}$ and $m_W$ are:
 \bea
\sin^2\theta_{eff} =0.23152\pm0.00022,\nonumber \\
m_W = 80.375\pm0.030 GeV
 \label{10a} 
 \eea 
 The fitted value of
$\sin^2\theta_{eff}$ is identical to the LEP+SLD average and the caution
on the error expressed in the previous section applies. The error of 30
MeV on
$m_W$  clearly sets up a goal for the direct measurement of $m_W$ at LEP2
and the Tevatron.

As a final comment we want to recall that the radiative corrections are
functions of $log(m_H)$. It is truly remarkable that the fitted value of
$log(m_H)$ is found to fall right into the very narrow  allowed window
around the value 2 specified by the lower limit from direct searches,
$m_H>77~GeV$, and the theoretical upper limit in the SM $m_H< 600-800~GeV$
(see sect.6). The fulfilment of this very stringent consistency check is a
beautiful argument in favour of a fundamental Higgs (or one with a
compositeness scale much above the weak scale).

\section{A More General Analysis of Electroweak Data }

We now discuss an update of the epsilon analysis~\cite{abc}$^,$\cite{abcnew}
which is a method to look at the data in a more general context than the
SM. The starting point is to isolate from the data that part which is due
to the purely weak radiative corrections. In fact the epsilon variables
are defined in such a way that they are zero in the approximation when
only effects from the SM at tree level plus pure QED and pure QCD
corrections are taken into account. This very simple version of improved
Born approximation is a good first approximation  according to the data
and is independent of $m_t$ and $m_H$. In fact the whole $m_t$ and $m_H$
dependence arises from weak loop corrections and therefore is only
contained in the epsilon variables. Thus the epsilons are extracted from
the data without need of specifying
$m_t$ and $m_H$. But their predicted value in the SM or in any extension
of it depend on $m_t$ and $m_H$. This is to be compared with the
competitor method based on the S, T, U variables~\cite{pes}$^,$\cite{abar}.
The latter cannot be obtained from the data without specifying
$m_t$ and $m_H$ because they are defined as deviations from the complete
SM prediction for specified $m_t$ and
$m_H$. Of course there are very many variables that vanish if pure weak
loop corrections are neglected, at least one for each relevant observable.
Thus for a useful definition we choose a set of representative observables
that are used to parametrize those hot spots of the radiative corrections
where new physics effects are most likely to show up. These sensitive weak
correction terms include vacuum polarization diagrams which being
potentially quadratically divergent are likely to contain all possible non
decoupling effects (like the quadratic top quark mass dependence in the
SM). There are three independent vacuum polarization contributions. In the
same spirit, one must add the $Z\rightarrow b \bar b$ vertex which also
includes a large top mass dependence. Thus altogether we consider four
defining observables: one asymmetry, for example
$A_{FB}^l$, (as representative of the set of measurements that lead to the
determination of
$\sin^2\theta_{eff}$), one width (the leptonic width
$\Gamma_l$ is particularly suitable because it is practically independent
of $\alpha_s$), $m_W$ and $R_b$. Here lepton universality has been taken
for granted, because the data show that it is verified within the present
accuracy. The four variables,
$\epsilon_1$, $\epsilon_2$, $\epsilon_3$ and $\epsilon_b$ are defined in
ref.~\cite{abc} in one to one correspondence with the set of observables 
$A^{FB}_l$, $\Gamma_l$,
$m_W$, and $R_b$. The definition is so chosen that the quadratic top mass
dependence is only present  in
$\epsilon_1$ and
$\epsilon_b$, while the
$m_t$ dependence of
$\epsilon_2$ and $\epsilon_3$ is logarithmic. The definition of
$\epsilon_1$ and $\epsilon_3$ is specified in terms of $A^{FB}_l$ and
$\Gamma_l$ only. Then adding $m_W$ or $R_b$ one obtains $\epsilon_2$ or
$\epsilon_b$. The values of the epsilons as obtained~\cite{abcnew},
following the specifications of ref.~\cite{abc}, from the defining
variables are shown in the first column of table 5.

\begin{table}[t]
\caption[]{Experimental values of the epsilons in the SM from
different sets of data. These values (in $10^{-3}$ units) are obtained for
$\alpha_s(m_Z) = 0.119\pm0.003$, $\alpha(m_Z) = 1/128.90\pm0.09$, the 
corresponding uncertainties being included in the quoted errors~\cite{abcnew}.}
\vspace{0.2cm}
\begin{center}
\footnotesize
\begin{tabular}{|l|l|l|l|l|}
\hline $\epsilon~~~10^3$  &Only def. quantities &All asymmetries &All High
Energy & All Data\\
\hline
$\epsilon_1~10^3$ &$4.0\pm1.2$ &$4.3\pm1.2$ &$4.1\pm1.2$  &$3.9\pm1.2$ \\
$\epsilon_2~10^3$ &$-8.3\pm2.3$ &$-9.1\pm2.1$ &$-9.3\pm2.2$  &$-9.4\pm2.2$
\\
$\epsilon_3~10^3$ &$2.9\pm1.9$ &$4.3\pm1.4$ &$4.1\pm1.4$  &$3.9\pm1.4$ \\
$\epsilon_b~10^3$ &$-3.2\pm2.3$ &$-3.3\pm2.3$ &$-3.9\pm2.1$ 
&$-3.9\pm2.1$  \\
\hline
\end{tabular}
\end{center}
\end{table}

To proceed further and include other measured observables in
the analysis we need to make some dynamical assumptions. The minimum
amount  of model dependence is introduced by including other purely
leptonic quantities at the Z pole such as $A_{\tau}$, $A_e$ (measured 
from the angular dependence of the $\tau$ polarization) and $A_{LR}$
(measured by SLD). For this step, one is simply assuming that the
different leptonic asymmetries are equivalent measurements of
$\sin^2\theta_{eff}$ (for an example of a peculiar model where this is not
true, see ref.~\cite{carLR}).  We add, as usual, the measure of
$A^{FB}_b$ because this observable is dominantly sensitive to the leptonic
vertex. We then use the combined value of $\sin^2\theta_{eff}$ obtained
from the whole set of asymmetries measured at LEP and SLC with the error
increased according to eq.(\ref{8}) and the related discussion. At this
stage the best values of the epsilons are shown in the second column of
table 5. In figs. 2-4  we report the 1$\sigma$ ellipses in the indicated
$\epsilon_i$-$\epsilon_j$ planes that correspond to this set of input
data. The status of
$\epsilon_b$ is shown in fig.5. The central value of $\epsilon_b$ is
shifted with respect to the SM as a consequence of the still imperfect
matching of $R_b$. In fig.5 we also give a graphical representation of the
uncertainties due to $\alpha(m_Z)$ and
$\alpha_s(m_Z)$.

\begin{figure}
\hglue 2.5cm
\epsfig{figure=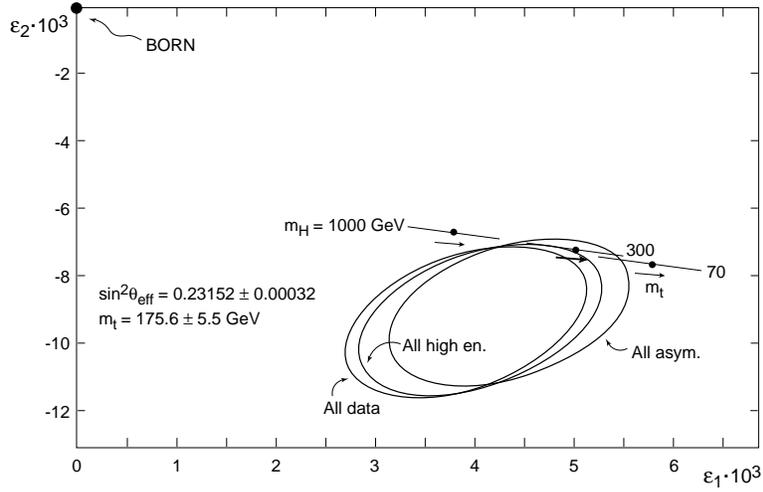,width=10cm}
\caption[]{Data vs theory in the $\epsilon_2$-$\epsilon_1$ plane. The origin point
corresponds to the "Born" approximation obtained from the SM at tree level
plus pure QED and pure QCD corrections. The predictions of the full SM
(also including the improvements of ref.~\cite{deg}) are shown for $m_H$ =
70, 300 and 1000 GeV and $m_t=175.6\pm5.5~GeV$ (a segment for each $m_H$ with
the arrow showing the direction of $m_t$ increasing from $-1\sigma$ to
$+1\sigma$). The three
$1-\sigma$ ellipses ($38\%$ probability contours) are obtained from a)
"All Asymm." :$\Gamma_l$, $m_W$ and
$\sin^2\theta_{eff}$ as obtained from the combined asymmetries (the value
and error used are shown); b) "All High En.": the same as in a) plus all
the hadronic variables at the Z; c) "All Data": the same as in b) plus the
low energy data.}
\end{figure}

\begin{figure}
\hglue 2.5cm
\epsfig{figure=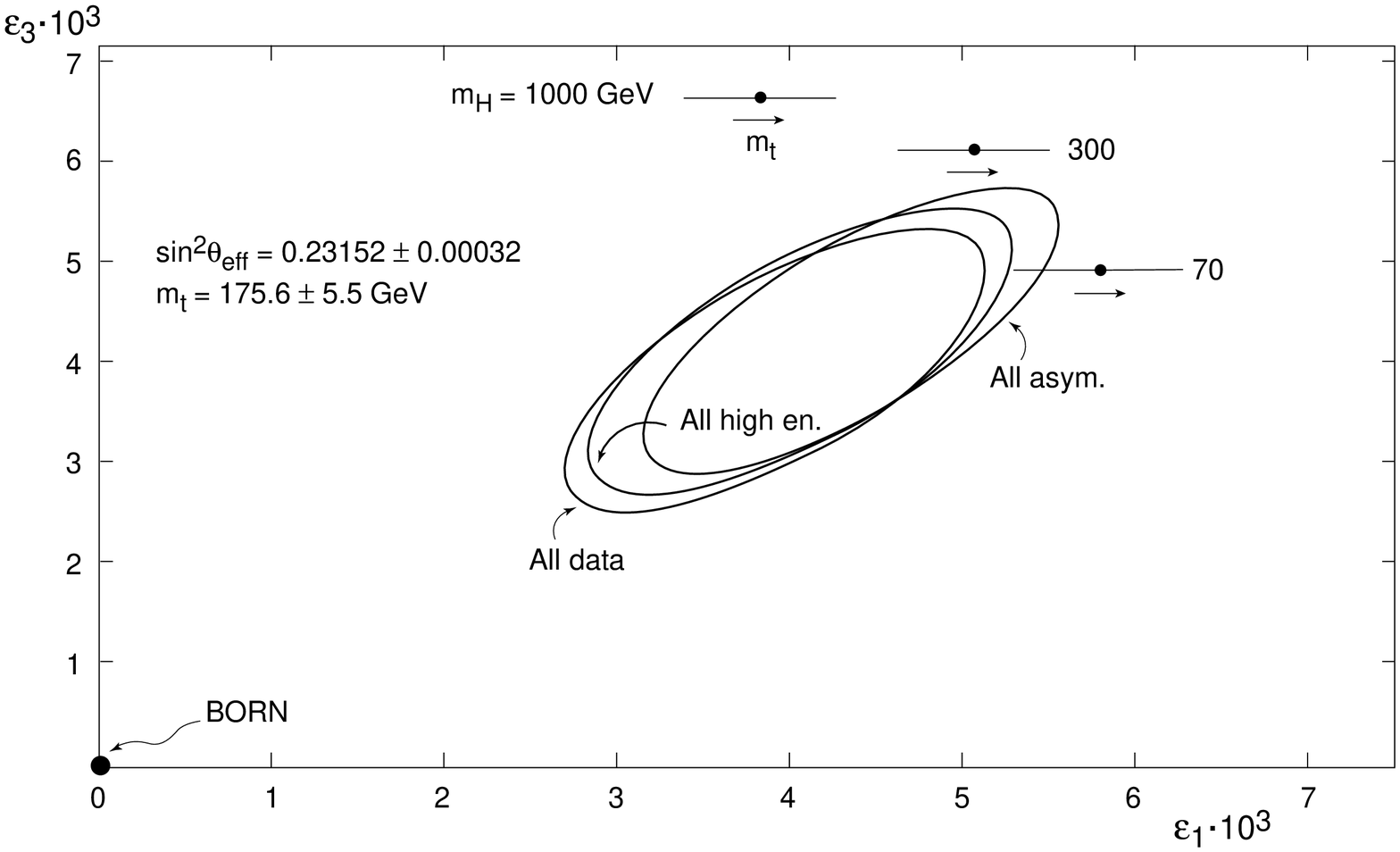,width=10cm}
\caption[]{Data vs theory in the $\epsilon_3$-$\epsilon_1$ plane (notations
 as in fig.2).}
\end{figure} 

\begin{figure}
\hglue 2.5cm
\epsfig{figure=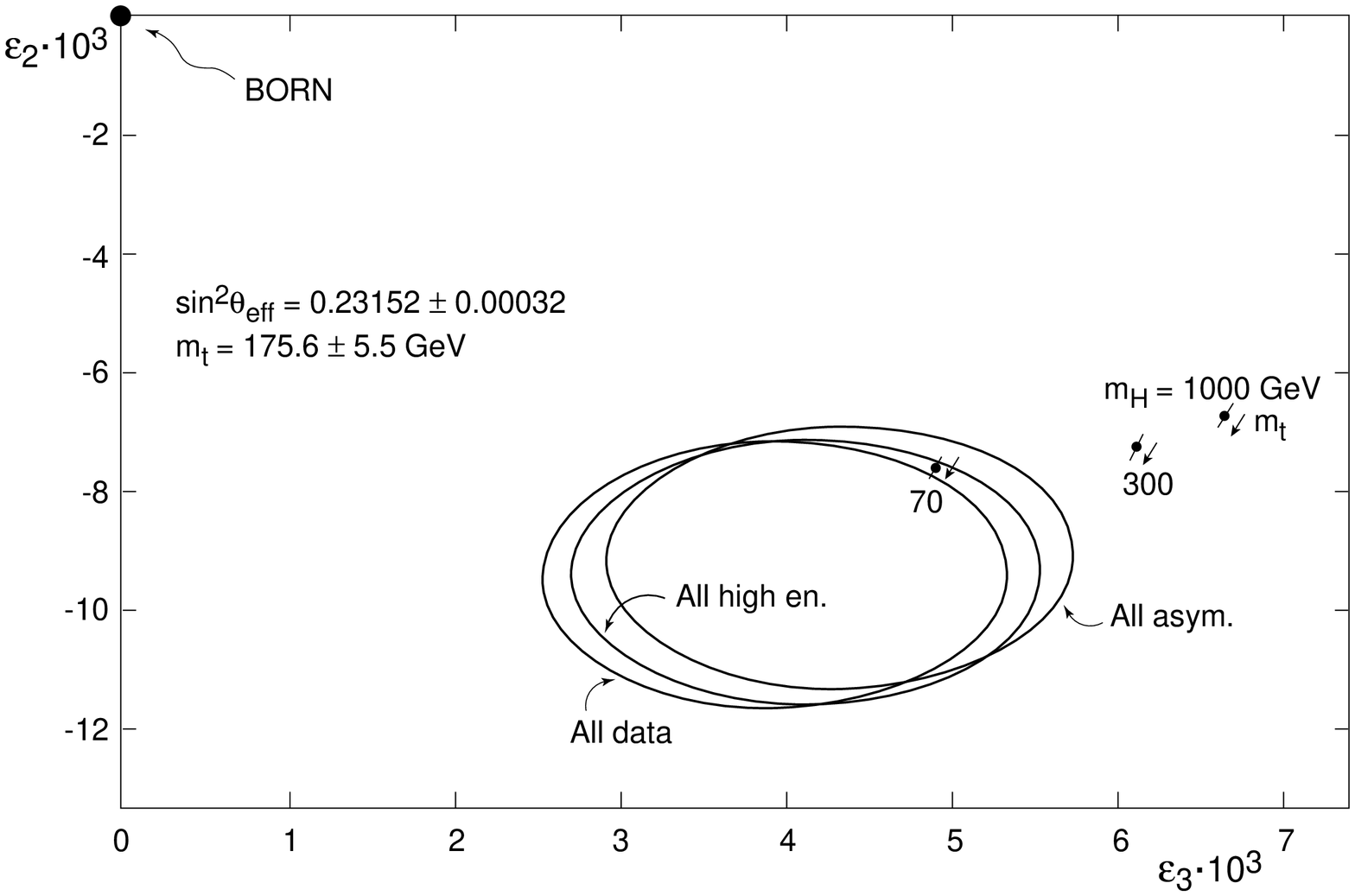,width=10cm}
\caption[]{Data vs theory in the $\epsilon_2$-$\epsilon_3$ plane (notations as in
fig.2).}
\end{figure}

\begin{figure}
\hglue 2.5cm
\epsfig{figure=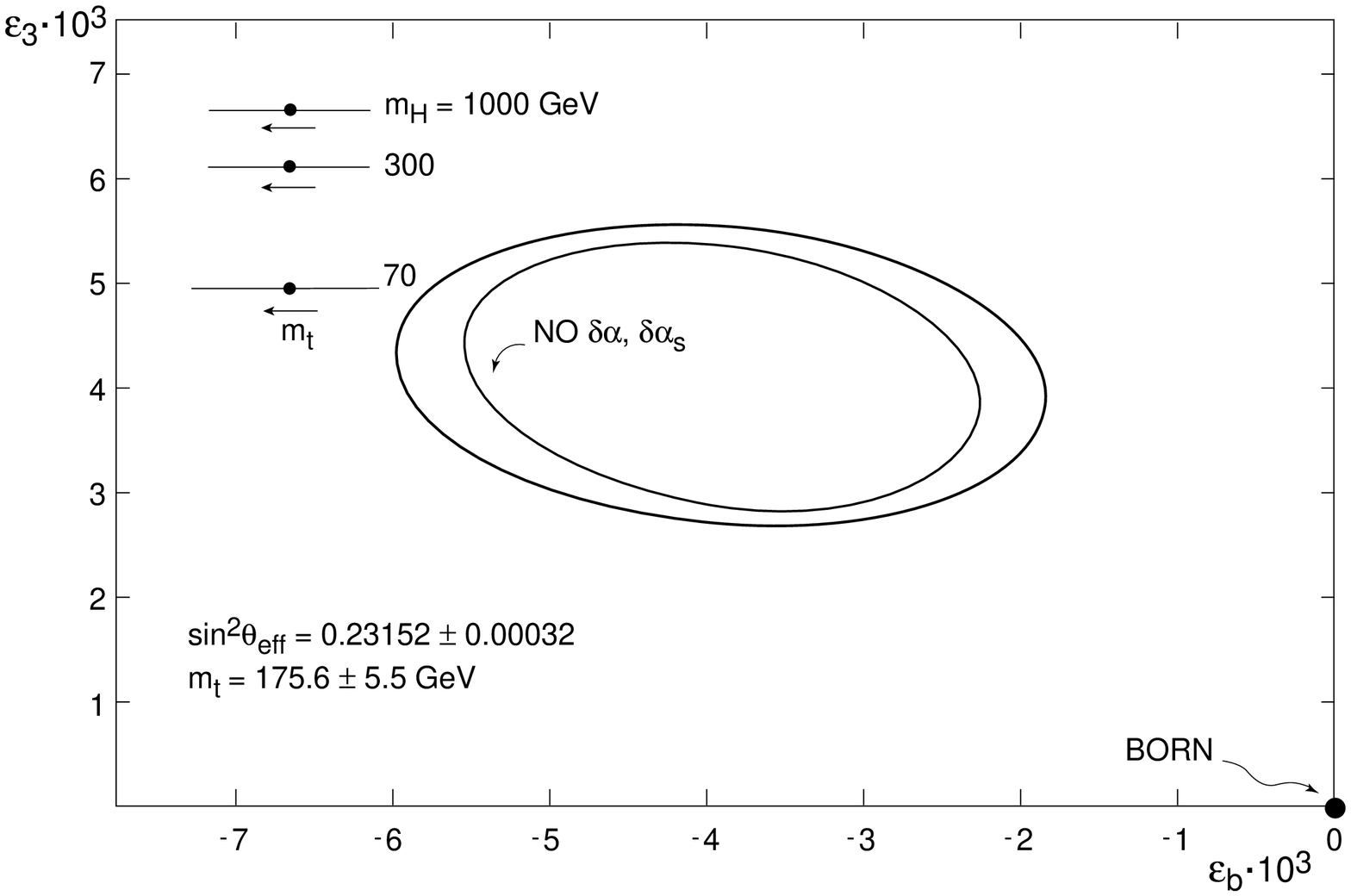,width=10cm}
\caption[]{Data vs theory in the $\epsilon_3$-$\epsilon_b$ plane (notations as in
fig.2, except that both ellipses refer to the case b)) The inner
$1-\sigma$ ellipse is without the errors induced by the uncertainties on
$\alpha(m_Z)$ and $\alpha_s(m_Z)$.}
\end{figure}

        All observables measured on the Z peak at LEP can be included in the
analysis provided that we assume that all deviations from the SM are only
contained in vacuum polarization diagrams (without demanding a truncation
of the $q^2$ dependence of the corresponding functions) and/or the
$Z\rightarrow b\bar b$  vertex. From a global fit of the data on $m_W$, 
$\Gamma_T$,  $R_h$, $\sigma_h$,  $R_b$ and
$\sin^2\theta_{eff}$ (for LEP data, we have taken the correlation matrix
for $\Gamma_T$,  $R_h$ and
$\sigma_h$ given by the LEP experiments~\cite{ew}, while we have
considered the additional information on $R_b$ and $\sin^2\theta_{eff}$ 
as independent) we obtain the values shown in the third column of table 5.
The comparison of theory and experiment at this stage is also shown in
figs. 2-5. 

         To include in our analysis lower energy observables as well, a stronger
hypothesis needs to be made:  vacuum polarization diagrams are allowed to
vary from the SM  only in their constant and first derivative terms in a
$q^2$ expansion~\cite{pes}$^-$\cite{abar}. In such a case, one can, for
example, add to the analysis the ratio
$R_\nu$ of neutral to charged current processes in deep inelastic neutrino
scattering on nuclei~\cite{33}, the "weak charge" $Q_W$  measured in atomic parity violation
experiments on Cs~\cite{34}  and the measurement of $g_V/g_A$ from
$\nu_\mu e$ scattering~\cite{35}. In this way one obtains  the global fit
given in the fourth column of table 5 and shown in figs. 2-5. With the
progress of LEP the low energy data, while important as a check that no
deviations from the expected
$q^2$ dependence arise, play a lesser role in the global fit. Note that
the present ambiguity on the value of $\delta\alpha^{-1}(m_Z) =\pm0.09$~\cite{alfaQED} corresponds to an uncertainty on $\epsilon_3$ (the other
epsilons are not much affected) given by $\Delta\epsilon_3~10^3 =\pm0.5$~\cite{abc}. Thus the theoretical error is still
comfortably less than the
experimental error. 

A number of interesting features are clearly visible from figs.2-5. First,
the good agreement with the SM (all the epsilons are within  $\lappeq1
\sigma$ from the SM value) and the evidence for weak corrections, measured
by the distance of the data from the improved Born approximation point
(based on tree level SM plus pure QED or QCD corrections). There is by now
a solid evidence for departures from the improved Born approximation where
all the epsilons vanish. In other words a clear evidence for the pure weak
radiative corrections has been obtained and LEP/SLC are now measuring the
various components of these radiative corrections. For example, some
authors~\cite{39} have studied the sensitivity of the data to a
particularly interesting subset of the weak radiative corrections, i.e.
the purely bosonic part. These terms arise from virtual exchange of gauge
bosons and Higgses. The result is that indeed the measurements are
sufficiently precise to require the presence of these contributions in
order to fit the data. Second, the general results of the SM fits are
reobtained from a different perspective. We see the preference for light
Higgs manifested by the tendency for
$\epsilon_3$ to be rather on the low side. Since $\epsilon_3$ is
practically independent of $m_t$, its low value demands $m_H$ small. If
the Higgs is light then the preferred value of
$m_t$ is somewhat lower than the Tevatron result (which in the epsilon
analysis is not included among the input data). This is because also the
value of $\epsilon_1\equiv \delta \rho$, which is determined by the
widths, in particular by the leptonic whidth, is somewhat low. In
particular
$\epsilon_1$ increases with $m_t$ and, at fixed $m_t$, decreases with
$m_H$, so that for small $m_H$ the low central value of $\epsilon_1$
pushes $m_t$ down. Note that also the central value of $\epsilon_2$ is on
the low side, because the experimental value of $m_W$ is a little bit too
large. Finally, we see that adding the hadronic quantities or the low
energy observables hardly makes a difference in the
$\epsilon_i$-$\epsilon_j$ plots with respect to the case with only the
leptonic variables being included (the ellipse denoted by "All Asymm.").
But, for example for the
$\epsilon_1$-$\epsilon_3$ plot, while the leptonic ellipse contains the
same information as one could obtain from a
$\sin^2\theta_{eff}$ vs $\Gamma_l$ plot, the content of the other two
ellipses is much larger because it shows that the hadronic as well as the
low energy quantities match the leptonic variables without need of any new
physics. Note that the experimental values of $\epsilon_1$ and
$\epsilon_3$ in the latter case also depend on the input value of
$\alpha_s$ given in eq.(\ref{9}).

\section{The Case for Physics beyond the Standard Model}

        Given the striking success of the SM why are we not satisfied with that
theory? Why not just find the Higgs particle, for completeness, and
declare that particle physics is closed? The main reason is that there are
strong conceptual indications for physics beyond the SM. 

        It is considered highly unplausible that the origin of the electro-weak
symmetry breaking can be explained by the standard Higgs mechanism,
without accompanying new phenomena. New physics should be manifest at
energies in the TeV domain. This conclusion follows fron an extrapolation
of the SM at very high energies. The computed behaviour of the
$SU(3)\otimes SU(2)\otimes U(1)$ couplings with energy clearly points
towards the unification of the electro-weak and strong forces (Grand
Unified Theories: GUTS) at scales of energy
$M_{GUT}\sim  10^{14}-10^{16}~ GeV$ which are close to the scale of
quantum gravity, $M_{Pl}\sim 10^{19}~ GeV$~\cite{qqi}$^,$\cite{bar}.  One can also imagine  a unified theory of all
interactions also including gravity (at present superstrings~\cite{ler} provide the best attempt at such a theory). Thus GUTS and the
realm of quantum gravity set a very distant energy horizon that modern
particle theory cannot anymore ignore. Can the SM without new physics be
valid up to such large energies? This appears unlikely because the
structure of the SM could not naturally explain the relative smallness of
the weak scale of mass, set by the Higgs mechanism at $m\sim
1/\sqrt{G_F}\sim  250~ GeV$  with $G_F$ being the Fermi coupling constant.
This so-called hierarchy problem~\cite{ssi} is related to the presence of fundamental scalar fields in the
theory with quadratic mass divergences and no protective extra symmetry at
m=0. For fermions, first, the divergences are logaritmic and, second, at
m=0 an additional symmetry, i.e. chiral  symmetry, is restored. Here, when
talking of divergences we are not worried of actual infinities. The theory
is renormalisable and finite once the dependence on the cut off is
absorbed in a redefinition of masses and couplings. Rather the hierarchy
problem is one of naturalness. If we consider the cut off as a
manifestation of new physics that will modify the theory at large energy
scales, then it is relevant to look at the dependence of physical
quantities on the cut off and to demand that no unexplained enormously
accurate cancellation arise. 

        According to the above argument the observed value of $m\sim 250~ GeV$ is
indicative of the existence of new physics nearby. There are two main
possibilities. Either there exist fundamental scalar Higgses but the
theory is stabilised by supersymmetry, the boson-fermion symmetry that
would downgrade the degree of divergence from quadratic to logarithmic.
For approximate supersymmetry the cut off is replaced by the splitting
between the normal particles and their supersymmetric partners. Then
naturalness demands that this splitting (times the size of the weak gauge
coupling) is of the order of the weak scale of mass, i.e. the separation
within supermultiplets should be of the order of no more than a few TeV.
In this case the masses of most supersymmetric partners of the known
particles, a very large managerie of states, would fall, at least in part,
in the discovery reach of the LHC. There are consistent, fully formulated
field theories constructed on the basis of this idea, the simplest one
being the MSSM \cite{43}. Note that all normal observed states are those
whose masses are forbidden in the limit of exact
$SU(2)\otimes U(1)$. Instead for all SUSY partners the masses are allowed
in that limit. Thus when supersymmetry is broken in the TeV range but
$SU(2)\otimes U(1)$ is intact only s-partners take mass while all normal
particles remain massless. Only at the lower weak scale the masses of
ordinary particles are generated. Thus a simple criterium exists to
understand the difference between particles and s-particles.

        The other main avenue is compositeness of some sort. The Higgs boson is
not elementary but either a bound state of fermions or a condensate, due
to a new strong force, much stronger than the usual strong interactions,
responsible for the attraction. A plethora of new "hadrons", bound by the
new strong force would  exist in the LHC range. A serious problem for this
idea is that nobody sofar has been  able to build up a realistic model
along these lines, but that could eventually be explained by a lack of
ingenuity on the theorists side. The most appealing examples are
technicolor theories \cite{30}$^-$\cite{31}. These models where inspired by
the breaking of chiral symmetry in massless QCD induced by quark
condensates. In the case of the electroweak breaking new heavy
techniquarks must be introduced and the scale analogous to $\Lambda_{QCD}$
must be about three orders of magnitude larger. The presence of such a
large force relatively nearby has a strong tendency to clash with the
results of the electroweak precision tests \cite{32}.

        The hierarchy problem is certainly not the only conceptual problem of the
SM. There are many more: the proliferation of parameters, the mysterious
pattern of fermion masses and so on. But while most of these problems can
be postponed to the final theory that will take over at very large
energies, of order $M_{GUT}$ or
$M_{Pl}$, the hierarchy problem arises from the unstability of the low
energy theory and requires a solution at relatively low energies. 

A supersymmetric extension of the SM provides a way out which is well
defined, computable and that preserves all virtues of the SM.  The
necessary SUSY breaking can be introduced through soft terms that do not
spoil the good convergence properties of the theory. Precisely those terms
arise from supergravity when it is spontaneoulsly broken in a hidden
sector \cite{yyi}. But alternative mechanisms of SUSY breaking are also
being considered~\cite{gauge}.  In the most familiar approach SUSY is broken in a hidden
sector and the scale of SUSY breaking is very large of order
$\Lambda\sim\sqrt{G^{-1/2}_F M_P}$  where
$M_P$ is the Planck mass. But since the hidden sector only communicates
with the visible sector through gravitational interactions the splitting
of the SUSY multiplets is much smaller, in the TeV energy domain, and the
Goldstino is practically decoupled. In an alternative scenario the (not so
much) hidden sector is connected to the visible one by ordinary gauge
interactions. As these are much stronger than the gravitational
interactions, $\Lambda$ can be much smaller, as low as 10-100 TeV. It
follows that the Goldstino is very light in these models (with mass of
order or below 1 eV typically) and is the lightest, stable SUSY particle,
but its couplings are observably large. The radiative decay of the
lightest neutralino into the Goldstino leads to detectable photons. The
signature of photons comes out naturally in this SUSY breaking pattern:
with respect to the MSSM, in the gauge mediated model there are typically
more photons and less missing energy. Gravitational and gauge mediation
are extreme alternatives: a spectrum of intermediate cases is conceivable.
The main appeal of gauge mediated models is a better protection against
flavour changing neutral currents. In the gravitational version even if we
accept that gravity leads to degenerate scalar masses at a scale near
$M_{Pl}$ the running of the masses down to the weak scale can generate
mixing induced by the large masses of the third generation fermions
\cite{bar}.

At present the most direct phenomenological evidence in favour of
supersymmetry is obtained from the unification of couplings in GUTS.
Precise LEP data on $\alpha_s(m_Z)$ and $\sin^2{\theta_W}$ confirm what
was already known with less accuracy: standard one-scale GUTS fail in
predicting $\sin^2{\theta_W}$ given
$\alpha_s(m_Z)$ (and $\alpha(m_Z)$) while SUSY GUTS \cite{zzi} are in
agreement with the present, very precise, experimental results. According
to the recent analysis of ref.\cite{aaii}, if one starts from the known
values of
$\sin^2{\theta_W}$ and $\alpha(m_Z)$, one finds for $\alpha_s(m_Z)$ the
results:
\bea
\alpha_s(m_Z) &=& 0.073\pm 0.002 ~~~~~(\rm{Standard~GUTS})\nonumber \\  
\alpha_s(m_Z)& =& 0.129\pm0.010~~~~~(\rm{SUSY~GUTS})
\label{24}
\eea 
to be compared with the world average experimental value
$\alpha_s(m_Z)$ =0.119(4).

Some experimental hints for new physics beyond the SM come from the sky
and from cosmology. I refer to solar and atmospheric neutrinos, dark
matter and baryogenesis. It seems to me that by now it is difficult to
imagine that the neutrino anomalies can all disappear or be explained away
\cite{aru}$^,$\cite{tot}. Probably they are a manifestation of neutrino
oscillations, hence neutrino masses. Massive neutrinos are natural in most
GUT's. The see-saw mechanism
\cite{ssm} provides an elegant explanation of the smallness of neutrino
masses, inversely proportional to the large scale of L violation. Minimal
SU(5) is disfavoured by neutrino masses, because there is no $\nu_R$ and
B-L is conserved by gauge interactions, while SO(10) provides a very
natural context for them \cite{qqi}. A number of different observations
show that most of the matter in the universe is non luminous and that both
cold and hot dark matter forms are needed \cite{tur}. Neutrinos with a
mass of a few eV's could provide the hot dark matter (particles that are
relativistic when they drop out of equilibrium). But neutrinos cannot be
the totality of dark matter because they are too faintly interacting and
have no enough clumping at galactic distances. Cold dark matter particles
do not exist in the SM. Plausible candidates are axions or neutralinos. In
this respect SUSY in the MSSM version scores a good point while other SUSY
options, like the gauge mediated models and the R-parity violating
versions, do not offer such a simple possibility. Baryogenesis is
interesting because it could occur at the weak scale \cite{rub} but not in
the SM. For baryogenesis one needs the three famous Sakharov conditions
\cite{sak}: B violation, CP violation and no termal equilibrium. In
principle these conditions could be verified in the SM. B is violated by
instantons when kT is of the order of the weak scale (but B-L is
conserved). CP is violated by the CKM phase and out of equilibrium
conditions could be verified during the electroweak phase transition. So
the conditions for baryogenesis appear superficially to be present for it
to occur at the weak scale in the SM. However, a more quantitative
analysis \cite{rev}$^,$\cite{cw1} shows that baryogenesis is not possible in
the SM because there is not enough CP violation and the phase transition
is not sufficiently strong first order, unless
$m_H<40~GeV$, which is by now excluded by LEP. Certainly baryogenesis
could also occur  below the GUT scale, after inflation. But only that part
with
$|B-L|>0$ would survive and not be erased at the weak scale by instanton
effects. Thus baryogenesis at $kT\sim 10^{12}-10^{15}~GeV$ needs B-L
violation at some stage like for $m_\nu$. The two effects could be related
if baryogenesis arises from leptogenesis \cite{lg} then converted into
baryogenesis by instantons. While baryogenesis at a large energy scale is
thus not excluded it is interesting that recent studies have shown that
baryogenesis at the weak scale could be possible in the MSSM \cite{cw2}.
In fact, in this model there are additional sources of CP violations and
the bound on $m_H$ is modified by a sufficient amount by the presence of
scalars with large couplings to the Higgs sector, typically the s-top.
What is required is that $m_H\lappeq 80-100~GeV$ (in the LEP2 range!), a
s-top not heavier than the top quark and, preferentially, a small
$\tan{\beta}$.

\section{The Search for the Higgs}

The SM works with remarkable accuracy. But the experimental foundation of
the SM is not completed if the electroweak symmetry breaking mechanism is
not experimentally established. Experiments must decide what is true: the
SM Higgs or Higgs plus SUSY or new strong forces and Higgs compositeness. 

The theoretical limits on the Higgs mass play an important role in the
planning of the experimental strategy.  The large experimental value of
$m_t$ has important implications on
$m_H$ both in the minimal SM \cite{zziii}$^-$\cite{bbiiii} and in its
minimal supersymmetric extension\cite{cciiii}$^,$\cite{ddiiii}. 

        It is well known\cite{zziii}$^-$\cite{bbiiii} that in the SM with only one
Higgs doublet a lower limit on
$m_H$ can be derived from the requirement of vacuum stability. The limit
is a function of $m_t$ and of the energy scale $\Lambda$ where the model
breaks down and new physics appears. Similarly an upper bound on $m_H$
(with mild dependence on $m_t$) is obtained \cite{eeiiii} from the
requirement that up to the scale $\Lambda$ no Landau pole appears. If one
demands vacuum stability up to a very large scale,of the order of
$M_{GUT}$ or
$M_{Pl}$ then the resulting bound on
$m_H$ in the SM with only one Higgs doublet is given by \cite{aaiiii}:
\begin{equation} m_H(GeV) > 138 + 2.1 \left[ m_t - 175.6 \right] -
3.0~\frac{\alpha_s(m_Z) - 0.119}{0.004}~.
\label{25}
\end{equation} In fact one can show that the discovery of a Higgs particle
at LEP2, or $m_H\lappeq 100~GeV$, would imply that the SM breaks down at a
scale
$\Lambda$ of the order of a few TeV. Of course, the limit is only valid in
the SM with one doublet of Higgses. It is enough to add a second doublet
to avoid the lower limit. The upper limit on the Higgs mass in the SM is
important for assessing the chances of success of the LHC as an
accelerator designed to solve the Higgs problem. The upper limit
\cite{eeiiii} has been recently reevaluated \cite{hr}. For $m_t\sim
175~GeV$ one finds
$m_H\lappeq 180~GeV$ for $\Lambda\sim M_{GUT}-M_{Pl}$ and $m_H\lappeq
0.5-0.8~TeV$ for $\Lambda\sim 1~TeV$.

A particularly important example of theory where the bound is violated is
the MSSM, which we now discuss.  As is well known \cite{43}, in the MSSM
there are two Higgs doublets, which implies three neutral physical Higgs
particles and a pair of charged Higgses. The lightest neutral Higgs,
called $h$, should be lighter than
$m_Z$ at tree-level approximation. However, radiative corrections
\cite{ffiiii} increase the $h$ mass by a term proportional to $m^4_t$ and 
logarithmically dependent on the stop mass . Once the radiative
corrections are taken into account the $h$ mass still remains rather
small: for $m_t$ = 174~GeV one finds the limit (for all values of tg
$\beta)~m_h < 130$~GeV
\cite{ddiiii}. Actually there are reasons to expect that $m_h$ is well
below the bound. In fact, if $h_t$ is large at the GUT scale, which is
suggested by the large observed value ot $m_t$ and by a natural onsetting
of the electroweak symmetry breaking induced by $m_t$, then at low energy
a fixed point is reached in the evolution of $m_t$. The fixed point
corresponds to $m_t \sim 195 \sin\beta$~GeV (a good approximate relation
for tg $\beta = v_{up}/v_{down} < 10$). If the fixed point situation is
realized, then $m_h$ is considerably below the bound, $m_h\lappeq 100~GeV$
\cite{ddiiii}.

In conclusion, for $m_t \sim 175$~GeV, we have seen that, on the one hand,
if a Higgs is found at LEP the SM cannot be valid up to $M_{Pl}$. On the
other hand, if a Higgs is found at LEP, then the MSSM has good chances,
because this model would be excluded for $m_h > 130$~GeV.

\section{New Physics at HERA?}

\subsection{ Introduction}
 The HERA experiments H1~\cite{H1} and
ZEUS~\cite{ZEUS}, recently updated in ref.~\cite{LP}, have  reported an
excess of deep-inelastic $e^+p$ scattering events at large values of
$Q^2\gappeq1.5 \times 10^4$ GeV$^2$, in a domain not previously explored
by other experiments. The total $e^+p$ integrated luminosity was of 14.2
+9.5 = 23.7 pb$^{-1}$, at H1 and of  20.1+13.4 = 33.5 pb$^{-1}$ at ZEUS.
The first figure refers to the data before the '97 run
\cite{H1}$^,$\cite{ZEUS}, while the second one refers to part of the
continuing '97 run, whose results were presented at the LP'97 Symposium
\cite{LP}. Both experiments collected in the past about $1~pb^{-1}$ each
with an
$e^-$ beam. A very schematic description of the situation is as follows. At
$Q^2\gappeq15~10^4$ in the neutral current channel (NC), H1 observes 12+6
= 18 events while about 5+3 = 8 were expected and ZEUS observes 12 + 6 =
18 events with about 9 + 6 =15 expected. In the charged current channel
(CC), in the same range of $Q^2$, H1 observes 4+2 = 6 events while about
1.8+1.2 = 3 were expected and ZEUS observes 3 + 2 = 5 events with about
1.2 + 0.8 = 2 expected. The distribution of the first H1 data suggested a
resonance in the NC channel. In the interval
$187.5<M<212.5~GeV$, which corresponds to $x\simeq 0.4$, and $y>0.4$, H1
in total finds 7 + 1 = 8 events with about 1 + 0.5 = 1.5 expected. But in
correspondence of the H1 peak ZEUS observes a total of 3 events, just
about the number of expected events. In the domain $x>0.55$ and $y>0.25$
ZEUS observes 3 + 2 events with about 1.2 + 0.8 = 2 expected. But in the
same domain H1 observes only 1 event in total more or less as expected. 

We see that with new statistics the evidence for the signal remain meager.
The bad features of the original data did not improve. First, there is a
problem of rates. With more integrated luminosity than for H1, ZEUS sees
about the same number of events in both the NC and CC channels. Second, H1
is suggestive of a resonance (although the evidence is now less than it
was) while ZEUS indicates a large $x$ continuum (here also the new data
are not more encouraging).  The difference could in part, but apparently
not completely \cite{Dr-Ber}, be due to the different methods of mass
reconstruction used by the two experiments, or to fluctuations in the
event characteristics. Of course, at this stage, due to the limited
statistics, one cannot exclude the possibility that the whole effect is a
statistical fluctuation. All these issues will hopefully be clarified by
the continuation of data taking. Meanwhile, it is important to explore
possible interpretations of the signal, in particular with the aim of
identifying additional signatures that might eventually be able to
discriminate between different explanations of the reported excess.

\subsection{Structure Functions}

Since the observed excess is with respect to the SM expectation based on
the QCD-improved parton model, the first question is whether the effect
could be explained by some inadequacy of the conventional analysis without
invoking new physics beyond the SM. In the somewhat analogous case of the
apparent excess of jet production at large transverse energy $E_T$
recently observed by the CDF collaboration at the Tevatron \cite{CDF}, it
has been argued \cite{CTEQ} that a substantial decrease in the discrepancy
can be obtained by modifying the gluon parton density at large values of
$x$ where it has not been measured directly. New results \cite{E756} on
large $p_T$ photons appear to cast doubts on this explanation because
these data support the old gluon density and not the newly proposed one.
In the HERA case, a similar explanation appears impossible, at least for
the H1 data. Here quark densities are involved and they are well known at
the same $x$ but smaller $Q^2$ \cite{tung}$^,$\cite{rock}, and indeed the
theory fits the data well there. Since the QCD evolution is believed to be
safe in the relevant region of $x$, the proposed strategy is to have a new
component in the quark densities at very large $x$, beyond the measured
region, and small $Q^2$ which is driven at smaller $x$  by the evolution
and contributes to HERA when $Q^2$ is sufficiently large \cite{tung}.
However it turns out that a large enough effect is only conceivable at
very large $x$, $x\gappeq 0.75$, which is too large even for ZEUS. The
compatibility with the Tevatron is also an important constraint. This is
because $ep$ scattering is linear in the quark densities, while $p \bar p$
is quadratic, so that a factor of 1.5-2 at HERA implies a large effect
also at the Tevatron. In addition, many possibilities including intrinsic
charm \cite{charmint} (unless
$\bar c \not = c$ at the relevant $x$ values \cite{Thomas}) are excluded
from the HERA data in the CC channel \cite{Babu2}. More in general, if
only one type of density is modified, then  in the CC channel one obtains
too large an effect in the $\bar u$ and d cases and no effect at all in
the $\bar d$ and u cases \cite{Babu2}. In conclusion, it is a fact that
nobody sofar was able to even roughly fit the data. This possibility is to
be kept in mind if eventually the data will drift towards the SM and only
a small excess at particularly large
$x$ and
$Q^2$ is left in NC channel.

\subsection{Contact Terms} Still considering the possibility that the
observed excess is a non-resonant continuum, a rather general approach in
terms of new physics is to interpret the HERA excess as due to an
effective four-fermion ${\bar e} e {\bar q} q$ contact
interaction \cite{Eichten83} with a scale $\Lambda$ of order a few TeV. It
is interesting that a similar contact term of the
${\bar q} q {\bar q} q$ type, with a scale of exactly the same order of
magnitude, could also reproduce the CDF excess in jet production at large
$E_T$ \cite{CDF}. (Note, however, that this interpretation is not
strengthened by more recent data on the dijet angular
distribution \cite{CDFangle}). One has studied in detail
\cite{ALTARELLI97}$^,$\cite{CT} vector contact terms of the general form
\beq
\Delta L=\frac{4\pi\eta_{ij}}{(\Lambda^\eta_{ij})^2} \; \bar
e_i\gamma^{\mu}e_i \; \bar q_j\gamma_{\mu}q_j
\label{0a}.
\eeq with $i,j=L,R$ and $\eta$ a $\pm$ sign. Strong limits on these contact
terms are provided by LEP2 \cite{LEP2} (LEP1 limits also have been
considered but are less constraining
\cite{LEP1}), Tevatron \cite{CT-CDF} and atomic parity violation (APV)
experiments \cite{34}. The constraints are even more stringent for scalar
or tensor contact terms. APV limits essentially exclude all relevant
$A_eV_q$ component. The CDF limits on Drell-Yan production are
particularly constraining. Data exist both for electron and muon pairs up
to pair masses of about 500 GeV and show a remarkable $e-\mu$ universality
and agreement with the SM. New LEP limits  (especially from LEP2) have
been presented \cite{LEP2}. In general it would be possible to obtain a
reasonably good fit of the HERA data, consistent with the APV and the LEP
limits, if one could skip the CDF limits \cite{barg}.  But, for example,
a parity conserving combination
$({\bar e}_L\gamma^\mu e_L)(  {\bar u}_R\gamma_\mu u_R) + ( {\bar
e}_R\gamma^\mu e_R )( {\bar u}_L\gamma_\mu u_L)$ with
$\Lambda^+_{LR}=\Lambda^+_{RL}\sim 4$ TeV  still leads to a marginal fit to
the HERA data and is compatible with all existing
limits \cite{barg}$^,$\cite{dib} (see fig.6 \cite{dib1}). Because we expect
contact terms to satisfy $SU(2)\bigotimes U(1)$, because they reflect
physics at large energy scales, the above phenomenological form is to be
modified into $\bar L_L \gamma_{\mu} L_L (\bar u_R \gamma^{\mu} u_R +\bar
d_R \gamma^{\mu} d_R) + \bar e_R
\gamma_{\mu} e_R \bar Q_L \gamma^{\mu} Q_L)$, where L and Q are doublets
\cite{car1}. This form is both gauge invariant and parity conserving.
Here one has taken into account the requirement that contact terms
corresponding to CC are too constrained to appear. More sophisticated fits
have also been performed \cite{barg}.

\begin{figure}
\hglue 2.5cm
\epsfig{figure=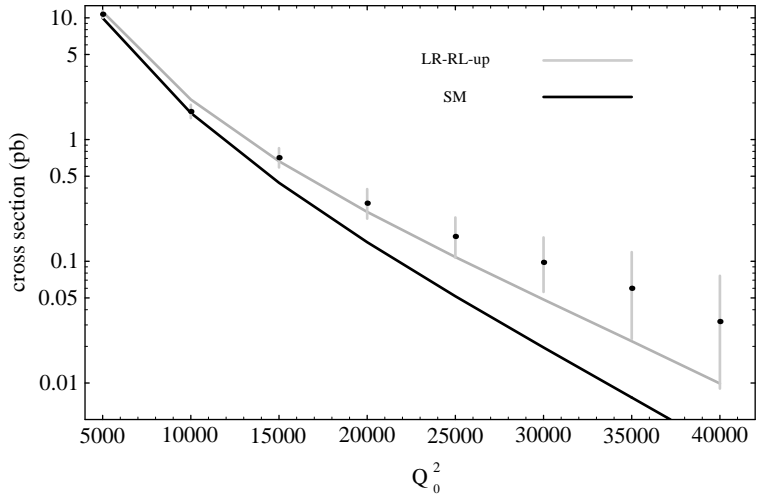,width=10cm}
\caption[]{Example of a fit to the HERA data presented at LP'97 from a LR+RL
contact term with only u-quarks
\cite{dib1} .}
\end{figure}
        
In conclusion, contact terms are severely constrained but not excluded.
The problem of generating the phenomenologically required contact terms
from some form of new physics at larger energies is far from trivial
~\cite{car1,stru}. Note also that contact terms require values of
$g^2/\Lambda^2 \sim 4\pi/(3-4~{\rm TeV})^2$, which would imply a very
strong nearby interaction. Indeed for $g^2$ of the order of the
$SU(3)\bigotimes SU(2)
\bigotimes U(1)$ couplings, $\Lambda$ would fall below 1~TeV, where the
contact term description is inadequate. We recall that the effects of
contact terms should be present in both the
$e^+$ and the
$e^-$ cases with comparable intensity. Definitely contact terms cannot
produce a CC signal \cite{alta2}, as we shall see, and no events with
isolated muons and missing energy.

\subsection{Leptoquarks} I now focus on the possibility of a resonance
with $e^+ q$ quantum numbers, namely a leptoquark
\cite{ALTARELLI97,Buch86,lqlimits,HERAlq,Bluemlein96,mont,donc}, of mass
$M\sim 190-210~GeV$, according to H1. The most obvious possibility is that
the production at HERA occurs from valence $u$ or $d$ quarks, since
otherwise the coupling would need to be quite larger, and more difficult
to reconcile with existing limits. However production from the sea is also
considered. Assuming an
$S$-wave state, one may have either a scalar or a vector leptoquark. I
only consider here the first option, because vector leptoquarks are more
difficult to reconcile with their apparent absence at the Tevatron. The
coupling
$\lambda$ for a scalar
$\phi$ is defined by
$\lambda \phi {\bar e}_L q_R$ or $\lambda \phi {\bar e}_R q_L$, The
corresponding width is given by
$\Gamma=\lambda^2M_{\phi}/16\pi$, and the production cross section on a
free quark is given in lowest order by $ \sigma \; = \; \frac{\pi}{4 s} \,
\lambda^2\;$.

Including also the new '97 run results, the combined H1 and ZEUS data,
interpreted in terms of scalar leptoquarks lead to the following list of
couplings \cite{Kunszt97,ALTARELLI97,Mlm}:
\beq e^+u\rightarrow\lambda \sqrt{B}\sim 0.017-0.025;~~~
e^+d\rightarrow\lambda \sqrt{B}\sim 0.025-0.033;~~~ e^+s\rightarrow\lambda
\sqrt{B}\sim 0.15-0.25\label{1a}
\eeq where B is the branching ratio into the e-q mode. By s the strange
sea is meant. For comparison note that the electric charge is
$e=\sqrt{4\pi \alpha}\sim0.3$. Production via
$e^+\bar u$ or
$e^+\bar d$ is excluded by the fact that in these cases the production in
$e^-u$ or
$e^- d$ would be so copious that it should have shown up in the small
luminosity already collected in the $e^-p$ mode. The estimate of $\lambda$
in the strange sea case is merely indicative due to the large
uncertainties on the value of the small sea densities at the relatively
large values relevant to the HERA data. The width is in all cases narrow:
for $B\sim 1/2$ we have
$\Gamma
\sim 4-16~MeV$ for valence and $350-1000~MeV$ for sea densities.

It is important to notice that improved data from the CDF and
D0\cite{E756} on one side and from APV \cite{34} and LEP \cite{LEP2} on
the other considerably reduce the window for leptoquarks. Consistency with
the Tevatron, where scalar leptoquarks are produced via model-independent
(and $\lambda$-independent) QCD processes with potentially large rates,
demands a value of B sizeably smaller than 1. In fact, the most recent NLO
estimates of the squark and leptoquark production cross
sections \cite{Spira96,Spira97} allow to estimate that at 200~GeV
approximately 6--7 events with $e^+e^-jj$ final states should be present
in the combined CDF and D0 data sets. For $B = 1$ the CDF limit is
210~GeV, the latest D0 limit is 225~GeV at 95\%CL. The combined CDF+D0
limit is 240~GeV at 95\%CL \cite{E756}.  We see that for consistency one
should impose:
\beq B \lappeq 0.5-0.7\\ \label{6a}
\eeq Finally, the case of a 200~GeV vector leptoquark  is most likely
totally ruled out by the Tevatron data, since the production rate can be
as much as a factor of 10 larger than that of scalar leptoquarks.

There are also lower limits on B, different for production off valence or
sea quarks, so that only a definite window for B is left in all cases. For
production off valence the limit arises from APV \cite{34}, while for the
sea case it is obtained from recent LEP2 data~\cite{LEP2}.

One obtains a limit from APV because the s-channel echange amplitude for a
leptoquark is equivalent at low energies to an $(\bar e q)(\bar q e)$
contact term with amplitude proportional to $\lambda^2/M^2$. After Fierz
rearrangement a component on the relevant APV amplitude $A_eV_q$ is
generated, hence the limit on $\lambda$. The results are~\cite{alta2}
\beq e^+u\rightarrow\lambda\lappeq 0.058;~~~ e^+d\rightarrow\lambda\lappeq
0.055\label{3a}
\eeq The above limits are for $M=200~GeV$ (they scale in proportion to M)
and are obtained from the quoted error on the new APV measurement on Cs.
This error being mainly theoretical, one could perhaps take a more
conservative attitude and somewhat relax the limit. Comparing with the
values for
$\lambda
\sqrt{B}$ indicated by HERA, given in eq.(\ref{1a}), one obtains lower
limits on B:
\beq e^+u\rightarrow B\gappeq 0.1-0.2;~~~~~ e^+d\rightarrow B\gappeq
0.2-0.4\label{33a}
\eeq For production off the strange sea quark the upper limit on $\lambda$
is obtained from LEP2~\cite{LEP2}, in that the t-channel exchange of the
leptoquark contributes to the process $e^+e^-\rightarrow s \bar s$
(similar limits for valence quarks are not sufficiently constraining,
because the values of $\lambda$ required by HERA are considerably
smaller). Recently new results have been presented by ALEPH, DELPHI and
OPAL~\cite{LEP2}. The best limits are around $\lambda\lappeq 0.6-0.7$
This, given eq.(\ref{1a}), corresponds to  
\beq e^+s
\rightarrow B\gappeq 0.05-0.2\\
\label{5a}
\eeq Recalling the Tevatron upper limits on B, given in eq.(\ref{6a}),  we
see that only a definite window for B is left in all cases. 

Note that one given leptoquark cannot be present both in $e^+p$ and in
$e^-p$ (unless it is produced from strange quarks). 

\subsection{S-quarks with R-parity Violation}

I now consider specifically leptoquarks and SUSY
~\cite{Hewett,Kon,Dreiner,CR,HERARPV,ALTARELLI97}. In general, in SUSY one
could consider leptoquark models without R-parity violation. It is
sufficient to introduce together with scalar leptoquarks also the
associated spin-1/2 leptoquarkinos~\cite{Moha}. In this way one has not
to give up the possibility that neutralinos provide the necessary cold
dark matter in the universe. We find it more attractive to embed a
hypothetical leptoquark in the minimal supersymmetric extension of the
Standard Model~\cite{43} with violation of
$R$ parity~\cite{RPth}. The connection with the HERA events has been more
recently invoked in ref.~\cite{CR}. The corresponding superpotential can
be written in the form
\beq W_R \equiv \mu_i H L_i + \lambda_{ijk} L_i L_j E^c_k + \lambda'_{ijk}
L_i Q_j D^c_k + \lambda ''_{ijk} U^c_i D^c_j D^c_k ~,
\label{WR}
\eeq where $H, L_i, E^c_j, Q_k, (U,D)^c_l$ denote superfields for the
$Y= 1/2$ Higgs doublet, left-handed lepton doublets, lepton singlets,
left-handed quark doublets and quark singlets, respectively. The indices
$i,j,k$ label the three generations of quarks and leptons.  Furthermore,
we assume the absence of the $\lambda''$ couplings, so as to avoid rapid
baryon decay, and the $\lambda$ couplings are not directly relevant in the
following.

The squark production mechanisms permitted by the $\lambda'$ couplings in
(\ref{WR}) include $e^+ d$ collisions to form
${\tilde u}_L, {\tilde c}_L$ or $\tilde t_L$, which involve valence $d$
quarks, and various collisions of the types $e^+ {d_i}$ ($i=2,3$) or $e^+
{\bar u_i}$ ($i=1,2,3$) which involve sea quarks. A careful analysis leads
to the result that the only processes that survive after taking into
account existing low energy limits are
\beq e^+_Rd_R \rightarrow \tilde c_L;~~~ e^+_Rd_R \rightarrow \tilde t_L
;~~~ e^+_Rs_R \rightarrow \tilde t_L\\
\label{7a}
\eeq For example $e^+_Rd_R \rightarrow \tilde u_L$ is forbidden by data on
neutrinoless double beta decay which imply 
\cite{Hirsch}
\beq
\vert \lambda'_{111} \vert < 7 \times 10^{-3} \left( {m_{\tilde q} \over
200 ~\hbox{GeV}}\right)^2
\left( {m_{\tilde g} \over 1 ~\hbox{TeV}}\right)^{1 \over 2}~.
\label{betabeta}
\eeq where $m_{\tilde q}$ is the mass of the lighter of
${\tilde u}_L$ and ${\tilde d}_R$, and $m_{\tilde g}$ is the gluino mass.

It is interesting to note~\cite{Kon} that the left s-top could be a
superposition of two mass eigenstates
$\tilde t_1,\tilde t_2$, with a difference of mass that can be large as it
is proportional to $m_t$:
\beq
\tilde t_L = \cos\theta_t~\tilde t_1~+~\sin\theta_t~\tilde t_2\\
\label{8a}
\eeq where $\theta_t$ is the mixing angle. With $m_1 \sim 200~GeV$, $m_2
\sim 230~GeV$ and $\sin^2\theta_t \sim 2/3$ one can obtain a broad mass
distribution, more similar to the combined H1 and ZEUS data. (But with the
present data one has to swallow that H1 only observes $\tilde t_1$ while
ZEUS only sees $\tilde t_2$!). However, the presence of two light
leptoquarks makes the APV limit more stringent. In fact it becomes
\beq B<B_{\infty} [1 + \tan^2{\theta_t}\frac{m^2_1}{m^2_2}]\\
\label{9a}
\eeq Thus, for the above mass and mixing choices, the above quoted APV
limit $B_{\infty}$ must be relaxed invoking a larger theoretical
uncertainty on the Cs measurement.

Let us now discuss~\cite{ALTARELLI97} if it is reasonable to expect that
$\tilde c$ and
$\tilde t$ decay satisfy the bounds on the branching ratio B. A virtue of
s-quarks as leptoquark is that competition of R-violating and normal
decays ensures that in general $B<1$.

In the case of ${\tilde c}_L$, the most important possible decay modes are
the $R$-conserving ${\tilde c}_L\rightarrow c \chi^0_i$ ($i=1,..,4$) and
${\tilde c}_L\rightarrow s \chi^+_j$ ($j=1,2$), and the $R$-violating
${\tilde c}_L\rightarrow d e^+$, where
$\chi^0_i, \chi^+_j$ denote neutralinos and charginos, respectively. In
this case it has been shown that, if one assumes that
$m_{\chi^+_j} > 200$ GeV, then , in a sizeable domain of the parameter
space, the neutralino mode can be sufficiently suppressed so that
$B\sim 1/2$ as required (for example, the couplings of a higgsino-like
neutralino are suppressed by the small charm mass). 

In the case of ${\tilde t}_L$, it is interesting to notice that the
neutralino decay mode ${\tilde t}_L \to t \chi^0_i$ is kinematically closed
in a natural way. In order to obtain a large value of B in the case of
s-top production off d-quarks, in spite of the small value of $\lambda$,
it is sufficient to require that all charginos are heavy enough to forbid
the decay ${\tilde t}_L \to b \chi^+_j$. However, we do not really want to
obtain B too close to 1, so that in this case some amount of fine tuning
is required. Or, with charginos heavy, one could invoke other decay
channels as, for example, $\tilde t \rightarrow \tilde b W^+$
~\cite{kon2}. But the large splitting needed between $\tilde t$ and 
$\tilde b$ implies problems with the $\rho$-parameter of electroweak
precision tests, unless large mixings in both the s-top and s-bottom
sectors are involved and their values suitably chosen. To obtain $B\sim
1/2$ is more natural in the case of s-top production off s-quarks, because
of the larger value of $\lambda$, which is of the order of the gauge
couplings.

The interpretation of HERA events in terms of s-quarks with R-parity
violation requires a very peculiar family and flavour
structure~\cite{Barbieri97}. The flavour problem is that there are very
strong limits on products of couplings from absence of FCNC. The
unification problem is that nucleon stability poses even stronger limits
on products of
$\lambda$ couplings that differ by the exchange of quarks and leptons
which are treated on the same footing in GUTS. However it was found that
the unification problem can be solved and the required pattern can be
embedded in a grand unification framework~\cite{Barbieri97}. The already
intricated problem of the mysterious texture of masses and couplings is
however terribly enhanced in these scenarios.

\subsection{Charged Current Events}

We have mentioned that in the CC channel at $Q^2\gappeq15~10^4$ H1 and
ZEUS see a total of 11 events with 5 expected. The statistics is even more
limited than in the NC case, so one cannot at the moment derive any firm
conclusion on the existence and on the nature of an excess in that
channel. However, the presence or absence of a simultaneous CC signal is
extremely significant for the identification of the underlying new physics
(as it would also be the case for the result of a comparable run with an
$e^-$ beam, which however is further away in time). It is found that in
most of the cases the CC signal is not expected to arise
\cite{alta2,babu3,Care,kon2}. But if it is present at a comparable rate as
for the NC signal, the corresponding indications are very selective. In
fact the following results are found. Due to the existing limits on
charged current processes, it is not possible to find a set of contact
terms that satisfy $SU(2)\bigotimes U(1)$ invariance and lead to a
significant production of CC events. For leptoquarks, we recall that a
leptoquark with branching ratio equal to 1 in
$e^+q$ is excluded by the recent Tevatron limits. Therefore on one hand
some branching fraction in the CC channel is needed. On the other hand,
one finds that there is limited space for the possibility that a
leptoquark can generate a CC signal at HERA with one single parton quark
in the final state. This occurrence would indicate
$SU(2)\bigotimes U(1)$ violating couplings or couplings to a current
containing the charm quark. A few mechanisms for producing CC final states
from $\tilde c$ or $\tilde t$ have been proposed~\cite{alta2,Care}.
In all cases $\tilde c$ or $\tilde t$ lead to multiparton final states.
Since apparently the CC candidates are all with one single jet, some
strict requirements on the masses of the partecipating particles must be
imposed so that some partons are too soft to be visible while others
coelesce into a single visible jet. So, s-quarks with R-parity violating
decays could indeed produce CC events or events with charged leptons and
missing energy. But the observation of such events would make the model
much more constrained.

\section{Conclusion}

The HERA anomaly is an interesting feature that deserves further attention
and more experimental effort. But at the moment it does not represent a
convincing evidence of new physics. The same is true for the other few
possible discrepancies observed here and there in the data. The overall
picture remains in impressive agreement with the SM. Yet, for conceptual
reasons, we remain
confident that new physics will eventually appear
at the LHC if not before.

It is for me a pleasure to thank my Scientific Secretary Dr. Matteo
Cacciari who was of great help for me and also Prof. Albrecht Wagner and
Dr. Albert De Roeck for their excellent organisation of the Symposium and
their kind assistance in Hamburg.

\end{document}